\documentclass[10pt,A4paper,conference]{IEEEtran}
\usepackage[english]{babel}
\usepackage{amsfonts}
\usepackage{epsfig}
\usepackage{amsmath}
\usepackage{amssymb}
\usepackage{hyperref}
\usepackage{picins}

\begin{document}

\title{Generic Security Proof of Quantum Key Exchange using Squeezed States}

\author{\authorblockN{Karin Poels}
\authorblockA{Dep. of Math. and Comp. Science\\
Eindhoven University of Technology\\
Eindhoven, The Netherlands \\
Email: k.j.p.m.poels@tue.nl}
\and
\authorblockN{Pim Tuyls}
\authorblockA{Information Security Systems\\
Philips Research Eindhoven\\
Eindhoven, The Netherlands\\
Email: pim.tuyls@philips.com}
\and
\authorblockN{Berry Schoenmakers}
\authorblockA{Dep. of Math. and Comp. Science\\
Eindhoven University of Technology \\
Eindhoven, The Netherlands\\
Email: berry@win.tue.nl}}

\maketitle

\begin{abstract}
Recently, a Quantum Key Exchange protocol that uses squeezed states was presented by Gottesman and Preskill. In this paper we give a generic security proof for this protocol. The method used for this generic security proof is based on recent work by Christiandl, Renner and Ekert. 
\end{abstract}

\section{Introduction}
In a Quantum Key Exchange (QKE) protocol there are three parties; Alice and Bob who want to exchange a secret key and a malicious third party, Eve. Eve has access to unlimited quantum computational power and she can monitor but not alter all public communication between Alice and Bob. Alice and Bob can access a quantum communication channel (we assume that this channel is lossless) and an authenticated public channel. 

Following a certain QKE protocol, Alice and Bob transmit quantum states over their quantum communication channel and perform measurements on their respective quantum states. From the measurements they extract bit values. By communication over the authenticated public channel, Alice and Bob agree which bits
will be used for secret key generation. They estimate the bit error rate $\epsilon$ of these key bits with another round of public communication. Alice and Bob then apply information reconciliation and privacy amplification to the key bits so that they end with a shared bit string $K$. 

In \cite{renner}, a generic security proof is proposed by which the security of a wide class of QKE protocols is proved. It is based on the fact that privacy amplification is equally secure when an adversary's memory for data storage is quantum rather than classical (\cite{konig}). The generic security proof gives Alice and Bob a threshold $d$ for the bit error rate $\epsilon$. This means that if $\epsilon \leq d$, then $K$ is unconditionally secure. The generic security proof is applicable to QKE protocols that involve quantum systems with a finite number of degrees of freedom (in \cite{renner} and in \cite{shorpreskill} it was proved that BB84 is secure for $\epsilon\leq 11\%$). It does not immediately apply however to QKE protocols using quantum systems with an infinite number of degrees of freedom.

BB84 (\cite{bb84}) is a QKE protocol that works with two-dimensional quantum bits (qubits) which are encoded by single photons polarized in one out of two non-orthogonal bases. The unconditional security of BB84 is based on Alice's ability to prepare single photons, something that is still extremely difficult.

In \cite{gottpress}, a QKE protocol, which we denote by GP00, was presented that resembles BB84 but solves the problem of preparing single photons. GP00 works with infinite-dimensional squeezed states, which can be prepared by a laser. The squeezing parameter $r$ determines the amount of squeezing of a squeezed state. The more squeezing, the more difficult a squeezed state is to prepare, therefore we need a lower bound for $r$. In \cite{gottpress} it was proved that the protocol is secure if $\epsilon \leq 11\%$ hence if $r\geq 0.289$. In this paper we apply the generic security proof to GP00 and find the same thresholds. A generic security proof is advantageous because it can give more insight in the security of similar protocols. Further, we will discuss some remaining security issues of GP00. Finally, we pay some attention to transmitting more than one bit per squeezed state.

\section{Squeezed States}

Let $\alpha \in \mathbb{C}$ and $\zeta = r e^{i \phi}$ with $r \in \mathbb{R}$ and $\phi \in [0,\pi)$. To every $\zeta,\alpha$ there corresponds a squeezed state denoted by $|\zeta,\alpha\rangle$. It satisfies with equality the Heisenberg uncertainty relation with respect to the position and momentum operators $x$ and $p$ if and only if $\phi=0$ so $\zeta = r \in \mathbb{R}$. That is, $\sigma_x \sigma_p = \frac{1}{2}$.
In fact, if we measure the position or the momentum of the squeezed state $|\zeta=r,\alpha\rangle$, then the measured value $x$ or $p$ is distributed according to a Gaussian distribution with variance equal to respectively $\sigma_x^2 = \frac{1}{2}e^{2r}$ or $\sigma_p^2  =  \frac{1}{2}e^{-2r}$. We say that $r$ is the squeezing parameter and that $|r,\alpha\rangle$ is a minimum uncertainty squeezed state. 
\begin{figure}[h]
\begin{center}
\includegraphics[scale=0.55]{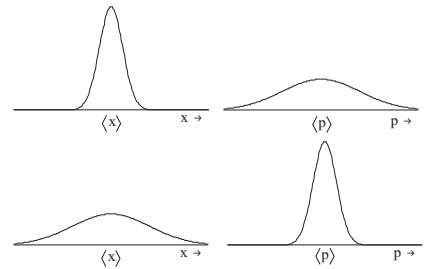}
\caption{Measurement probability distributions for measuring the position or the momentum of a squeezed state squeezed in a) $x$ and b) $p$. \label{fig:squeeze}}
\end{center}
\end{figure}
If $r<0$, then $\sigma_x^2<\sigma_p^2$ and the squeezed state is ``squeezed'' in $x$. If $r>0$, then the squeezed state is ``squeezed'' in $p$ (see Fig.\ \ref{fig:squeeze}). After a measurement of position value $x$ or momentum value $p$, the squeezed state collapses to respectively a position eigenstate $|x\rangle$ or a momentum eigenstate $|p\rangle$. 

\section{Bit Encoding and Decoding Scheme for GP00}\label{sec coding decoding}

First we fix 
$\hat{r}>0$. 
All squeezed states are squeezed with squeezing parameter $r=-\hat{r}$ (for squeezing in $x$) or $r=\hat{r}$ (for squeezing in $p$). 
\begin{figure}[h]
\begin{center}
\includegraphics[scale=0.8]{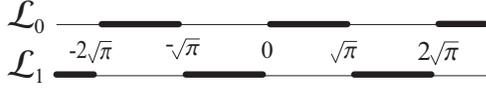}
\caption{Bit encoding intervals $\mathcal{L}_0$ and $\mathcal{L}_1$. \label{fig:encoding intervals}}
\end{center}
\end{figure}
We divide the real numbers into two sets of intervals $\mathcal{L}_0$ and $\mathcal{L}_1$ as in Fig.\ \ref{fig:encoding intervals}. Alice samples $a \in \mathbb{R}$ from the Gaussian $P_A(a)$ with mean $0$ and variance $\frac{1}{2}e^{2\hat{r}}$:  
\begin{eqnarray}
P_A(a)=\frac{1}{\sqrt{\pi e^{2\hat{r}}}}\exp{\left[-\frac{a^2}{e^{2\hat{r}}} \right]}.\label{Pa(a)}
\end{eqnarray}
If $a \in \mathcal{L}_0$, Alice extracts bit $0$, otherwise she extracts bit $1$. She prepares a squeezed state squeezed in $x$ or $p$ at random. 
If she squeezes in $x$, she sends to Bob 
\[ |-\hat{r},\alpha\rangle \mbox{\ with \ }  \left\{
\begin{array}{ll}
\langle x \rangle & = a' = a \sqrt{1-e^{-4 \hat{r}}} \\
\langle p \rangle & = 0
\end{array}
\right. \]
If she squeezes in $p$, she sends to Bob the squeezed state
\[ |\hat{r},\alpha\rangle \mbox{\ with \ }  \left\{
\begin{array}{ll}
\langle x \rangle & = 0 \\
\langle p \rangle & = a' = a \sqrt{1-e^{-4 \hat{r}}}
\end{array}
\right. \]
For every squeezed state Alice computes and announces $\phi = a \bmod{\sqrt{\pi}}$ where $0\leq \phi < \sqrt{\pi}$. Note that there exists an $n_a\in\mathbb{Z}$ such that $a=n_a\sqrt{\pi}+\phi$. Every value for $\phi$ should be equally likely because then $P(a\in\mathcal{L}_0|\phi)=P(a\in\mathcal{L}_0)=0.5$ such that $\phi$ leaks no information to Eve (we further discuss this in Section \ref{sec: randomization fi}).

For every squeezed state Bob decides at random to measure the position or the momentum. Suppose that the outcome of his measurement is $b$ and denote the difference of Alice's and Bob's value by $\delta= b - a$. Note that $b-\phi =n_a\sqrt{\pi}+\delta$. Bob extracts bit value $0$ if $b-\phi=n_a\sqrt{\pi}+\delta$ rounded to the nearest integer multiple of $\sqrt{\pi}$ is an even multiple of $\sqrt{\pi}$ and bit value $1$ otherwise.

\begin{figure}[h]
\begin{center}
\includegraphics[scale = 0.77]{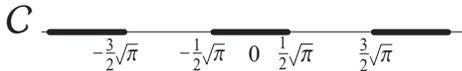}
\caption{Bit decoding interval $\mathcal{C}$. }\label{fig:decoding interval}
\end{center}
\end{figure}

If we define the decoding interval $\mathcal{C}$ as in Fig.\ \ref{fig:decoding interval}, then Alice and Bob find the same bit if $\delta = b-a \in \mathcal{C}$. This is because if $\delta\in\mathcal{C}$, then $n_a\sqrt{\pi}+\delta$ rounded to the nearest integer multiple of $\sqrt{\pi}$ is equal to $n_a\sqrt{\pi}$. They find different bits if $\delta = b-a  \notin \mathcal{C}$.  

\section{Bit Extraction Probabilities for GP00}

\begin{figure}[h]
\begin{center}
\includegraphics[scale = 0.48]{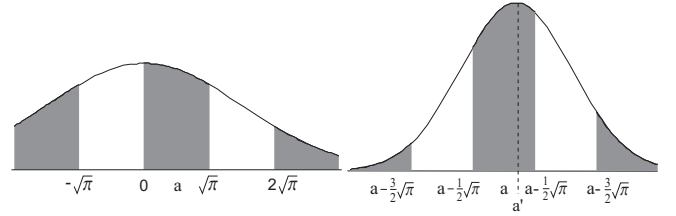}
\caption{Bit correct probability if a) different basis are used and $a=\frac{1}{2}\sqrt{\pi}$ and b) the same basis is used and $a'-a=\frac{1}{4}\sqrt{\pi}$ . }\label{fig: differnet basis}
\end{center}
\end{figure}

If Alice and Bob use different bases, then the value measured by Bob has a Gaussian distribution centered at $0$. This distribution is shown as the graph on the left in Fig.\ \ref{fig: differnet basis} for $a=\frac{1}{2}\sqrt{\pi}$. The marked area pictures the values of $b$ for which $b-a\in\mathcal{C}$ and represents the probability that Alice and Bob find the same bit. This probability equals $0.5$ if $a=\frac{1}{2}\sqrt{\pi}$. In fact, this probability is maximal if $a=2n\sqrt{\pi}$, is equal to $0.5$ if $a = (2n+\frac{1}{2})\sqrt{\pi}$ and is minimal if $a=(2n+1)\sqrt{\pi}$. This means that if all values for $\phi$ are equally likely, then the bit extracted by Bob is on average random. The corresponding cases in the protocol can therefore be discarded.
If Alice and Bob use the same basis, then the probability that they find the same bit is dependent on the distance between $a$, the value from which Alice extracts her bit, and $a'$, the mean value of Bob's squeezed state. This is illustrated in the graph on the right in Fig.\ \ref{fig: differnet  basis} for $a'-a=\frac{1}{4}\sqrt{\pi}$. We denote the average probability that Alice and Bob find the same bit, given that they use the same basis, by $1-\epsilon_s$. For illustration, this probability is $0.89$ if $\hat{r}=0.289$ and if $\hat{r}\rightarrow \infty$ it will approach $1$.
Note that even if Alice and Bob use the same basis then, in contrast to BB84, they find the same bit with probability smaller than $1$. This means that because of using squeezed states, additional quantum noise is introduced ($\epsilon_s$).

\section{The protocol GP00}\label{sec: protocol}
We give the description of the protocol. At the end of the protocol, just before information reconciliation and privacy amplification, Alice and Bob each have an $n$-bit string respectively $X$ and $Y$. After information reconciliation and privacy amplification they have a shared secure key $K$ of length $k<n$.
\begin{enumerate}
\item Alice prepares approximately $4n$ squeezed states. For every squeezed state she decides to squeeze it in $x$ or in $p$ at random. She prepares the squeezed states according to the encoding scheme described in the previous sections. For every squeezed state she extracts a bit value. She sends the squeezed states to Bob.
\item For each squeezed state, Bob decides to measure the position or the momentum at random.
\item Bob confirms having received the squeezed states. Alice and Bob announce which bases they used.
\item Alice and Bob discard the cases where they did not use the same basis. From the remaining approximately $4n/2=2n$ bits Alice chooses $n$ to serve as check bits and $n$ to serve as key bits. For the squeezed states corresponding to these check and key bits, Alice computes $\phi$. Alice sends all $\phi$'s and the chosen positions of the check and key bits to Bob such that Bob can  extract check and key bits from his measured values. Alice and Bob's resulting key bit strings are $X$ and $Y$.
\item Alice and Bob announce their check bits to estimate the bit error rate $\epsilon$.
\item If $\epsilon\leq 11\%$, then information reconciliation and privacy amplification follow such that Alice and Bob end with a shared secret key $K$.
\end{enumerate}
An important difference between GP00 and BB84 is that Bob needs additional information $\phi$ about the squeezed states to extract bit values from his measured values. Noise ($\epsilon$) is not only caused by the channel or by Eve, but also by the natural noise of squeezed states ($\epsilon_s$). If $\hat{r}\rightarrow \infty$, then $\epsilon_s \rightarrow 0$ and GP00 approaches the continuous version of $BB84$. 

\section{The Generic Security Proof}

The generic security proof \cite{renner} can be applied to a generic QKE protocol equivalent to an entanglement based protocol. A dealer prepares entangled states and sends one part of the entangled state to Alice and the other part to Bob. Let the measurements that Alice and Bob randomly apply to their received quantum states be the POVM's $\mathcal{F}$ and $\mathcal{G}$ and let the bit error rate of the bits extracted from the measurements be $\epsilon$. Let $\mathcal{R}$ be the set of all density operators $\rho$ (describing the quantum state of two systems) for which it holds that if $\rho$ is measured with respect to $\mathcal{F}\otimes \mathcal{F}$ or $\mathcal{G}\otimes \mathcal{G}$, then the two bits extracted from the measurement have bit error probability $\epsilon$. Thus $\mathcal{R}$ is the set of all possible density operators describing the mutual state of Alice and Bob, given that the bit error rate is equal to $\epsilon$. Let $\mathcal{Z}$ be a projective measurement on the density operator $\rho\in\mathcal{R}$ with outcome described by the random variable $Z$. Let $X$ and $Y$ be random variables such that Alice's and Bob's bit strings consist of $n$ realizations of these variables. The secret key rate $R$ is now given by (\cite{renner})
\begin{eqnarray}
R & = & H(X)-H(X|Y) - \arg_{\rho\in\mathcal{R}}{\max{H(Z)}}.\label{r3}
\end{eqnarray}
The rate might be improved by conditioning on additional information $W$, known only by Alice and Bob and gained during privacy amplification. The rate then becomes
\begin{eqnarray}
R & = & H(X|W)-H(X|Y) - \arg_{\rho\in\mathcal{R}}{\max{H(Z|W)}}. \ \ \ \  \label{r4}
\end{eqnarray}
The generic security proof consists in finding the maximum error rate $\epsilon$ such that $R$ is still positive and hence the extracted secret key $K$ is secure.

\section{Entanglement Based version of GP00}\label{section: entanglement based version of GP00}

To be able to apply the generic proof to GP00, we regard it as an entanglement based protocol. The entangled states prepared by the dealer (given in both position eigenstates and momentum eigenstates) are as follows (\cite{gottpress}) 
\begin{eqnarray*}
|\psi\rangle & = & \frac{1}{\sqrt{\pi}}\int \int \exp{\left[ -\frac{\Delta^2}{2}x_a^2\right]}\cdot \ \ \ \ \ \ \ \ \ \ \ \ \ \ \ \\ & & \exp{\left[-\frac{1}{2\Delta^2}\left(x_b-\sqrt{1-\Delta^4}x_a \right)^2\right]}|x_a,x_b\rangle  dx_b dx_a\\
 & = & \frac{1}{\sqrt{\pi}}\int \int \exp{\left[ -\frac{\Delta^2}{2}p_a^2 \right]} \cdot \\
  & & \exp{\left[ -\frac{1}{2\Delta^2}\left(p_b+\sqrt{1-\Delta^4}p_b \right)^2\right]}|p_a,p_b\rangle dp_b dp_a
\end{eqnarray*}
where $0<\Delta^2\leq 1$. For $\Delta^2 < 1$,  $|\psi\rangle$ is an entangled state. 

Alice and Bob both get a part of this entangled state. If Alice measures the position of her part, she measures position value $x_a$ with probability \[P_x(x_a)=\frac{\Delta}{\sqrt{\pi}}\exp{\left[-\Delta^2 x_a^2 \right]}.\]
By this measurement, she prepares for Bob the state
\[\frac{1}{(\pi \Delta^2)^{1/4}}\int \exp{\left[-\frac{1}{2\Delta^2}\left(x_b-\sqrt{1-\Delta^4}x_a \right)^2\right]}|x_b\rangle  dx_b,\]
which is a squeezed state squeezed in $x$ with mean position value $\sqrt{1-\Delta^4}x_a$ and mean momentum value $0$.
If Alice measures the momentum of her part, she measures value $p_a$ with probability $P_p(p_a)=P_x(p_a)$. By this measurement, she prepares for Bob a squeezed state squeezed in $p$ with mean momentum value $-\sqrt{1-\Delta^4}p_a$ and mean position $0$.

If we choose $\Delta^2 = e^{-2 \hat{r}}$, then the  entanglement based protocol is equivalent to GP00; from Eq. \ref{Pa(a)} we see that $P_x(x_a)=P_p(x_a)=P_A(x_a)$ and the squeezed states produced by Alice's measurements in the entanglement based version are equal to the squeezed states sent by Alice in GP00. Note that in the entanglement based version, the mean momentum value is $-\sqrt{1-\Delta^4}p_a$ rather than $ \sqrt{1-\Delta^4}p_a$. This means that Alice extracts a bit value and calculates $\phi$ from $-p_a$ rather than from $p_a$ if she measured the momentum and from $x_a$ if she measured the position. Alice and Bob find the same bit if $x_b-x_a\in\mathcal{C}$ or $p_b-(-p_a)=p_b+p_a \in\mathcal{C}$.

\section{Generic Security Proof of GP00}

Let $\epsilon$ be the bit error probability of the check bits. Let the density operator $\rho\in \mathcal{R}$; if of both parts of $\rho$ the position is measured or the momentum, then the probability that the extracted bits differ is equal to $\epsilon$. This can be formulated by
\begin{eqnarray}
\int_{-\infty}^{\infty}{\int_{x \in \mathcal{C}}{\langle x_a,x_a+x|\rho|x_a,x_a+x\rangle}dx}dx_a & = & 1-\epsilon  \ \ \ \ \ \label{a1} \\
\int_{-\infty}^{\infty}{\int_{x \in \mathcal{C}^c}{\langle x_a,x_a+x|\rho|x_a,x_a+x\rangle}dx}dx_a & = & \epsilon\label{a2} \\
\int_{-\infty}^{\infty}{\int_{p \in \mathcal{C}}{\langle p_a,-p_a+p|\rho|p_a,-p_a+p\rangle}dp}dp_a & = & 1-\epsilon\label{a3} \\
\int_{-\infty}^{\infty}{\int_{p \in \mathcal{C}^c}{\langle p_a,-p_a+p|\rho|p_a,-p_a+p\rangle}dp}dp_a & = & \epsilon\label{a4}
\end{eqnarray}
where e.g. $\langle x_a,x_a+x|\rho|x_a,x_a+x\rangle$ is the probability that Alice measures position value $x_a$ and Bob measures position value $x_a+x$.

As projective measurement $\mathcal{Z}$ we choose the continuous Bell measurement which is given by the projectors
$\{|\psi(x,p)\rangle\langle \psi(x,p)| | x, p \in \mathbb{R}\}$ with
\begin{eqnarray}
|\psi(x,p)\rangle & = & \int_{-\infty}^{\infty}{e^{ipx_a}|x_a,x_a+x\rangle dx_a} \nonumber\\
  & = & \int_{-\infty}^{\infty}{e^{ixp_a}|p_a,-p_a+p\rangle dp_a}.\nonumber
\end{eqnarray}
If we define \[\lambda_{xp} = \langle \psi(x,p)|\rho|\psi(x,p)\rangle,\]
then $\lambda_{xp}$ is the probability that if Alice and Bob both measure the position, then the difference of their outcomes is $x$ and if they both measure the momentum, then the sum of their outcomes equals $p$. If $x=x_b-x_a\in\mathcal{C}$ or $p=p_b+p_a \in\mathcal{C}$, then Alice and Bob extract the same bit and if $x\notin\mathcal{C}$ or $p\notin\mathcal{C}$ they extract different bits. 

We let $Z$, the random variable that describes the outcome of the projective measurement $\mathcal{Z}$, describe whether or not Alice and Bob will find the same bit, given that they both use the same basis. This leads to four different values (situations) for $Z$; 
we obtain the four probabilities corresponding to the four different values of $Z$ by grouping the probabilities $\lambda_{xp}$ in the following way
\[
\lambda_1  =  \int_{p\in\mathcal{C}}\int_{x\in\mathcal{C}}\lambda_{xp}dx dp \qquad \lambda_3  =  \int_{p\in\mathcal{C}}\int_{x\notin\mathcal{C}}\lambda_{xp}dx dp \]
\[ \lambda_2  =  \int_{p\notin\mathcal{C}}\int_{x\in\mathcal{C}}\lambda_{xp}dx dp \qquad \lambda_4  =  \int_{p\notin\mathcal{C}}\int_{x\notin\mathcal{C}}\lambda_{xp}dx dp.
\]
For illustration, $\lambda_2$ is the probability that if Alice and Bob both measure the position, they find the same bit and if they both measure the momentum, they find different bits. Eqs.\ (\ref{a1},\ref{a2},\ref{a3},\ref{a4}) can be rewritten as
\[\begin{array}{rclrcl}
\lambda_1 + \lambda_2 & = & 1-\epsilon \ \ \  & \lambda_3 + \lambda_4 & = & \epsilon \\
\lambda_1 + \lambda_3 & = & 1-\epsilon \ \ \ & \lambda_2 + \lambda_4 & = & \epsilon
\end{array}\]
With these relations, $\lambda_1,\lambda_2$ and $\lambda_3$ can be expressed in terms of $\lambda_4$. The entropy of the random variable $Z$ is given by 
\[H(Z)=-\sum_{i=1}^4{\lambda_i \log_2{\lambda_i}}\] and is maximized for $\lambda_4=\epsilon^2$ and then $H(Z)=2 h(\epsilon)$. The secret key bit rate becomes (Eq. \ref{r3}) 
\[R = 1-h(\epsilon)-2h(\epsilon)  =  1-3h(\epsilon).\] This rate is positive for $\epsilon\leq 6.1\%$.

We improve the rate by using the additional information $W=X+Y$ gained during privacy amplification. It holds that
\[\begin{array}{rcl}
H(Z|W) & = & \sum_{i\in\{0,1\}}P(W=i)H(Z|W=i) \\
 & = & (1-\epsilon) h\left(\frac{\lambda_1}{1-\epsilon}\right) + \epsilon h \left(\frac{\lambda_3}{\epsilon}\right) \\
 & = & \ H(Z) - h(\epsilon). 
\end{array}\]
The entropy $H(Z|W)$ is maximized for $\lambda_4 = \epsilon^2$ and then $H(Z|W)=h(\epsilon)$. The rate $R$ becomes (Eq. \ref{r4}) \[R  =  1-h(\epsilon)-h(\epsilon)  =  1-2h(\epsilon)\] which is positive for $\epsilon\leq 11\%$. This means that GP00 is secure if $\epsilon \leq 11 \%$. Because the noise generated by squeezed states ($\epsilon_s$) contributes to the total noise $\epsilon$, we have $\epsilon_s \leq \epsilon$. With calculations we find that $\epsilon_s\leq 11\%$ if $\hat{r}\geq 0.289$. This means that GP00 can only be secure if squeezed states are squeezed with squeezing parameter $\hat{r}\geq 0.289$.

\section{Randomization issue of \Large{$\phi$}.}\label{sec: randomization fi}

In GP00, Alice announces $\phi=a\bmod{\sqrt{\pi}}$. For unconditional security it has to hold that $P(a\in\mathcal{L}_0|\phi)=P(a\in\mathcal{L}_0)=0.5$ because then Alice can safely announce $\phi$ since it leaks no information to Eve. The probability $P(a\in\mathcal{L}_0|\phi)$ is maximal at $\phi=0$, equal to $0.5$ if $\phi=\frac{1}{2}\sqrt{\pi}$ and minimal at $\phi=\sqrt{\pi}$. For $\hat{r}=0.289$ we find e.g. that $P(a\in\mathcal{L}_0|\phi=0)=0.745$ which is rather high; it means that $\phi$ leaks a significant amount of information to Eve about the bit extracted by Alice. Study still has to be done in whether the generic security method allows $\phi$ to be non perfectly random. We considered three possible, alternative, solutions.  

One way to solve the problem, is to enlarge the lower bound for $\hat{r}$. For example, if $\hat{r}\geq1.5$, then $P(a\in\mathcal{L}_0|\phi=0)\approx0.5$ and no information leaks to Eve. Because $\hat{r}$ becomes considerably large, this solution is not favorable.
\begin{figure}[h]
\begin{center}
\includegraphics[scale=0.7]{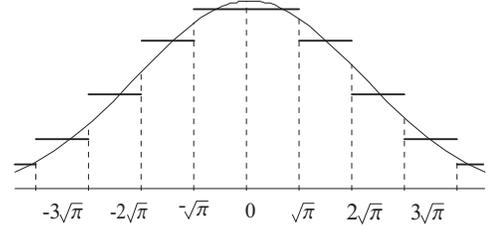}
\caption{Discrete approximation of $P_A(a)$.\label{fig: nieuwe verdeling}}
\end{center}
\end{figure}

A different solution that we considered is to make a discrete approximation of Alice's sampling distribution $P_A(a)$ as in Fig.\ \ref{fig: nieuwe verdeling}. It then holds that every value for $\phi$ is equally likely, the value $\phi$ does not leak information to Eve and if Bob or Eve measures the squeezed state in the incorrect basis, the extracted bit is on average random. 
If we use this discrete approximation to calculate the average error probability $\epsilon_s$ caused by squeezed states, then we find that for a given squeezing parameter $\hat{r}$, the value of $\epsilon_s$ increases (e.g. if $\hat{r}=0.289$ then $\epsilon_s=0.119$). We found that $\epsilon_s \leq 11\%$ if $\hat{r}\geq0.308$.
Although this discrete approximation seems to work, it is the case that the protocol resulting from the discrete approximation has no obvious entanglement based equivalent anymore such that the generic security proof \cite{renner} might not be applicable. We are still investigating the possibilities for this situation. In the following section we describe how we can transmit more than one bit per squeezed state, by making use of a similar discrete approximation of $P_A(a)$.  

It seems that for unconditional security, the bit encoding and/or decoding strategy of GP00 should be changed such that every value of $\phi$ becomes equally likely while the sampling distribution $P_A(a)$ and the squeezed states sent to Bob remain the same (this is the third solution we considered). By keeping $P_A(a)$ and the squeezed states sent to Bob the same, the resulting protocol has the entanglement based equivalent as described in Section \ref{section: entanglement based version of GP00}. An idea to do this is to choose $\phi=|a|\bmod{\sqrt{\pi}}$ instead of $\phi=a\bmod{\sqrt{\pi}}$. 

\section{Sending more bits per squeezed state}

We describe a method, based on a discrete approximation of $P_A(a)$, by which we can send $m$ bits per squeezed state. We show in more detail how to do this for $2$ bits, how the method works for more bits will follow easily from the $2$-bits case. The main difference with GP00 is the bit encoding and decoding scheme. We emphasize that unconditional security is not proven yet (see the comment on the discrete approximation solution in the previous section).

\begin{figure}[h]
\begin{center}
\includegraphics[scale=0.5]{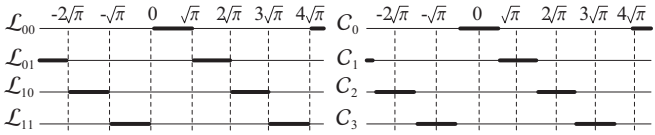} 
\caption{Encoding and decoding intervals for sending two bits.}\label{fig: encoding intervals for 2 bits}
\end{center}
\end{figure}

\begin{figure}[h]
\begin{center}
\includegraphics[scale=0.7]{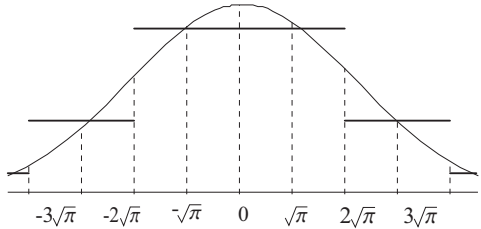} 
\caption{Discrete approximation of $P_a(a)$ if we send 2 bits per squeezed state.}\label{Discrete approximation of $P_a(a)$ if we send 2 bits per squeezed state.}
\end{center}
\end{figure}

Alice and Bob can extract $4$ different messages per squeezed state, which are given by
$m_0 = 00, m_1 = 01, m_2 = 10$ and $m_{2^2-1} = 11$. Alice samples $a$ from a discrete approximation of $P_A(a)$ such that all messages $m_0,\ldots m_3$ are equally likely. If $a\in\mathcal{L}_{00}$ then Alice extracts bits $m_0=00$, if $a\in\mathcal{L}_{01}$ then Alice extracts bits $m_1=01$ etc. where the encoding intervals $\mathcal{L}_{00},\ldots, \mathcal{L}_{11}$ are illustrated in the graph on the left in Fig.\ \ref{fig: encoding intervals for 2 bits}.
The discrete approximation of $P_A(a)$ that satisfies the constraint is illustrated in Fig.\ \ref{Discrete approximation of $P_a(a)$ if we send 2 bits per squeezed state.}.
With this discrete approximation, the value $\phi=a\bmod{\sqrt{\pi}}$ does not leak information to Eve and all messages $m_i$ are equally likely. Alice sends the same squeezed state to Bob, as she would send in GP00.
For every squeezed state Alice computes and announces $\phi = a \bmod{\sqrt{\pi}}$ where $0\leq \phi < \sqrt{\pi}$.
The procedure at Bob's side is also similar to that in GP00. For every squeezed state Bob decides at random to measure the position or the momentum. Suppose that the outcome of his measurement is $z\in\mathbb{R}$ and that Alice extracted the message $m_i$ where $i\in\{0,1,2,3\}$. Bob rounds $z-\phi$ to the nearest integer multiple of $\sqrt{\pi}$. If this integer multiple is equal to $j\bmod{2^2}$, then Bob extracts bit pair $m_j$. This means that Bob extracts bit pair $m_{(i+k)\bmod{4}}$ if $z-a\in\mathcal{C}_{k}$, for $k\in\{0,1,2,3\}$. The decoding intervals $\mathcal{C}_0,\ldots,\mathcal{C}_3$ are illustrated in the graph on the right in Fig.\ \ref{fig: encoding intervals for 2 bits}.

If Bob or Eve measures in the incorrect basis, then the message he or she extracts is random because of the discrete approximation used for $P_A(a)$. Suppose Alice and Bob use the same basis. Then, the lower bound for $\hat{r}$ seems to be comparable to that in the one-bit case. Research still has to be done for the exact value of the lower bound for $\hat{r}$. 

We can do similar reasoning for sending $3$ bits per squeezed states. In this case however, it seems that the lower bound for the squeezing parameter $\hat{r}$ increases. This is probably because the discrete approximation of $P_A(a)$ becomes too stretched. The lower bound seems to increase even more when we send more than $3$ bits per squeezed state.

\section{Concluding Remarks}
It has been shown how to use a generic security proof to prove the security of GP00, a QKE protocol that works with squeezed states. We studied a remaining weak point of the protocol and discussed some possible solutions. Elaborating on one of these possible solutions, we suggested a method to transmit more than one bit per squeezed state.


\begin{thebibliography}{1}

\bibitem{renner}
M. Christiandl, R. Renner \& A. Ekert, ``A Generic Security Proof for Quantum Key Distribution,''
2004, quant-ph/0402131.

\bibitem{konig} R. K\"onig, U. Maurer \& R. Renner; ``On the power of quantum memory,''2003, quant-ph/0305154.

\bibitem{bb84} C.H. Bennett \& G. Brassard; ``Quantum Cryptography; Public key distribution and coin tossing,'' 1984, Proceedings of IEEE International Conference on Computers, Systems and Signal Processing, 175-179, IEEE, New York.

\bibitem{gottpress} D. Gottesman \& J. Preskill, ``Secure quantum key exchange using squeezed states,'' 2000, quant-ph/0008046.

\bibitem{shorpreskill} P.W. Shor \& J. Preskill; ``Simple proof of security of the BB84 quantum key distribution protocol,'' 2000, Phys. Rev. Lett., 85(2):441-444, quant-ph/0003004

\end{thebibliography}
\end{document}